\documentclass[aps,prl,twocolumn,groupedaddress]{revtex4}
\begin{document}

%\preprint{}

\title{Analyzing the correlations for spin-$\frac{1}{2}$ particles and singlet pairs} 


\author{Edson de Faria} 
\affiliation{Instituto de Matem\'atica e Estat\'\i stica, USP, S\~ao Paulo, SP, Brazil}
\author{Charles Tresser} 
\affiliation{IBM, P.O.  Box 218, Yorktown Heights, NY 10598, U.S.A.}
\email[]{edson@ime.usp.br}
\email[]{charlestresser@yahoo.com}


\date{\today}

\begin{abstract}
We revisit the computation of correlations of spin projections onto unit vectors for spin-$\frac{1}{2}$ particles in Quantum Mechanics. We then choose one of the Boole inequalities that, as we recall, must be obeyed by collections of sequences of normalized spin-$\frac{1}{2}$ projections onto unit vectors that belong to, say, the plane that is orthogonal to the classical trajectory or to $\bf{R}^3$.  Next we define the concept of being \emph{$\bf{a}$-p} for $i$-indexed sequences of spin-$\frac{1}{2}$ particles: it turns out that a sequence of spin-$\frac{1}{2}$  particles is $\bf{a}$-p if and only if the particles behave as if they had indeed been prepared (in the usual sense) to have definite spins +1 or -1 along the axis $\bf{a}$, the sequence of signs being pre-determined.  The $\bf{a}$-p property is formulated in a way abstract enough to let one generalize it and we define the concept of being $\bf{a}$-p and $\bf{b}$-p.  However, using our chosen Boole inequality we prove that no particle sequence can be both $\bf{a}$-p and $\bf{b}$-p when  $\bf{a}$ and $\bf{b}$ are not co-linear.  We next consider two modifications of Quantum Mechanics: 

%\noindent
$\quad$- The first one by assuming both \emph{Weak Realism} (close to what is called \emph{Macroscopic Counterfactual Definiteness} 
by Leggett), constructed as the weakest form of Realism needed to develop any main theorem in Bell's Theory,  
and \emph{Locality}.
 
%\noindent  
$\quad$- The second one by assuming \emph{MCFD} and the \emph{Effect After Cause Principle} (or \emph{EACP}), an hypothesis that turns out to be none else than (relativistic) Causality extended to cover as well the observable values that make sense only if one assumes some form of Realism, (\emph{e.g.,} Weak Realism). 

\smallskip
%\noindent 
In both cases we arrive at the conclusion that in the EPRB setting where successive pairs of spin-$\frac{1}{2}$ particles are prepared in the so-called \emph{singlet state}, one of these particles sequences is both $\bf{a}$-p and  $\bf{b}$-p (for some $\bf{a}$ and $\bf{b}$ that are not co-linear) and $\bf{a}$ and $\bf{b}$ are the opposite of the vectors respectively used to project the spin of the other particle and to get the Realism-based alternate value for a projection.  \emph{Thus the usual discussions of Bell's inequalities under such hypotheses do not even make physical sense}. 
\end{abstract}

%\keywords{}

\maketitle

%\section{}
%\subsection{}
%\subsubsection{}
%
%
%
%
\section{Introduction}\label{sec:Introduction}
In this note we mostly consider ($i$-indexed) sequences of spin-$\frac{1}{2}$ particles and sequences of \emph{EPRB pairs}, \emph{i.e.,} pairs of spin-$\frac{1}{2}$ particles to be defined hereafter and named that way in reference to an exposition and reformulation using spin-$\frac{1}{2}$ particles by Bohm \cite{Bohm} of a paper \cite{EPR} often attributed to Einstein, Podolsky, and Rosen.  This paper was written by Podolsky and published without Einstein's \emph{imprimatur}  \cite{Jammer74}, \cite{Jammer85}, \cite{FineShaky}.  Einstein had his own way to tell what he considered as being the main content of his joint work with his co-authors, work that he considered as somewhat oddly reported in \cite{EPR} (see, \emph{e.g.,} \cite{EinsteinIdeasAndOpinions} and \cite{Schilpp}).  EPRB pairs are \emph{entangled} pairs of spin-$\frac{1}{2}$ particles such that the spin part of the wave function is the \emph{singlet state} 
\begin{equation}\label{Singlet}
\Psi(x_1,x_2)=\frac{1}{\sqrt{2}}(| +\rangle _A\otimes| -\rangle_B-| -\rangle_A\otimes|
+\rangle_B)\,.
\end{equation}
Following Stapp \cite{Stapp1971}, we insist on neutrality of the  \emph{EPRB particles} (\emph{i.e.,} members of EPRB pairs) so that the reader can think of preparations and measurements being performed with the best known and quite picturesque Stern-Gerlach magnets.  The singlet state is an example of \emph{entanglement} because one can prove that the sum of tensor products in (\ref{Singlet}) cannot be rewritten as a single tensor product of one particle states.  We even have here an example of \emph{maximal entanglement} since all summands in (\ref{Singlet}) have identical statistical weights. Under suitable hypotheses that we will review later on in this paper, singlet states can be proven to be subject to inequalities (re-)discovered by Bell \cite{Bell}.  Bell not only rediscovered an inequality of a kind studied by Boole in a different context \cite{Boole1854}, \cite{Boole1862} (as pointed out by Itamar Pitowsky: see \cite{Pitowsky2001} and references therein) but also formulated how \emph{Quantum Mechanics} (or \emph{QM}) should be \emph{augmented} (by the adjunction,\emph{e.g.,} of new hypotheses and/or extra postulates) to let one associate a large enough number of spin projections to a singlet state for an inequality of Boole's type and more precisely of Bell's type to apply.  

To avoid confusion, we will call \emph{Bell's Inequality} any inequality of the form considered by Boole and Bell when one can conclude from the quantities being compared by said inequality that \emph{Realism} (to be defined precisely later on in this paper) \emph{must be} invoked in order to have all the needed normalized spins' projections.  Then we say that we are in \emph{Bell's Realm}.  Otherwise speaking, in Bell's Realm, besides the observable values that can be simultaneously observed  we must also have some observable's values that can only be inferred to exist and only have value (that cannot be known) if one assumes some form of Realism. Thus in Bell's Realm there might be instances when there are no observed values although we will not meet such cases when all values being considered only exist because of some form of Realism.

On the other hand, \emph{Boole's Inequalities} are what one gets with QM (without any Realism) or in macroscopic physics when Realism is usually not even questioned: we then say that we are in \emph{Boole's Realm} (otherwise speaking, in Boole's Realm we only consider observables' values that can be (and indeed are!) simultaneously measured). 

\medskip
Many Boole's Inequalities can be written in (non augmented) QM although few are ever used (some are indeed quite trivial, \emph{e.g.,} the sum of the probabilities of an event and it contrary cannot exceed one).  These inequalities were devised by Boole in the context of studying data about collections of macroscopic objects to check that indexed collections of data supposed to come from similarly indexed samples were not mixed up (be it by accident or
on pupose \cite{Boole1854}, \cite{Boole1862}).   

To the contrary, Bell's Realm only refers to the augmentations of Quantum Mechanics, and in particular, the augmentations involving Realism.  Whether Realism cannot or can be deduced from Locality (as some authors have claimed) is orthogonal to the discussion that we are concerned with here.  

\medskip
We will discuss below augmentations of QM to the extent needed for this paper before we apply our Boole's Realm result on the non-existence of what we call \emph{``$\bf{a}$-p and $\bf{b}$-p particles"} (defined below) to Bell's Realm: we will conclude that Bell's Inequality have no physical content in Bell's Realm (even of course if one consider Bell's Realm as physically meaningful).  But first we shall deal with Boole's Realm, and more precisely we shall exhibit a Boole Inequality and its use in the context of ``QM as usual".  

\medskip
As we shall see, it turns out that the Boole's Inequality that is \emph{formally} (and formally only because of Realm considerations) \emph{the very inequality used by Bell in \cite{Bell}} because of Realm consideration) also has some interesting content when used in Boole's Realm.  This Boole and Bell (formal) 
inequality applies to any triple $(g,\,f,\, h )$ of sequences with values in $\{-1,\,+1\}$ similarly indexed, and reads:

\begin{equation}\label{BooleBell}
 |\,\langle{f},\, {g}\rangle -\langle{f}, \, {h}\rangle\,| +  \langle{g}, \, {h}\rangle\;\leq\;1\ ,
\end{equation}

\noindent
where the correlations means as usual that we are dealing with a typical sequence so that the average values of the products all converge to the values to be expected (and that can sometimes be better grasped in terms of probabilities ${\rm Prob}([Event(1)]_i=[Event(2)]_i)$ of equalities of some similarly indexed events).  Since for any two sequences of elements in $\{-1,\,+1\}$, say $\{f_i\}_{i\in \bf{N} }$ and  $\{g_i\}_{i\in \bf{N}}$, one has the relation:
\begin{equation}\label{BooleBell1}
{\rm Prob}(f_i=g_i)\, \equiv\,\frac{1+\langle f,\,g\rangle}{2}\,, 
\end{equation}
one can as well write the Boole and Bell Inequality (\ref{BooleBell}) in terms of the probabilities of coincidence, thus:
 \begin{equation}\label{BooleBellProb}
|\, {\rm Prob}(f_i=g_i) - {\rm Prob}(f_i=h_i) | \leq 1-   {\rm Prob}(g_i=h_i) .
\end{equation}

So far we have used the set $\bf{N}$ of natural integers as index set. It turns out that the index set that we will generally denote as $\bf{I}$ will typically be finite except perhaps for theory (when one should consider both finite and infinite index sets, although finite size corrections are often well documented).  Whether $\bf{I}$ is finite or $\bf{I}\,=\, \bf{N}$, one relies on standard approximation and convergence properties to make sense of the formulas, to write dowm the conditions of validity etc.  

%%%%%%%%%%%%%   Section 2   %%%%%%%%%%%
%%%%%%%%%%%%%   Section 2   %%%%%%%%%%%
%%%%%%%%%%%%%   Section 2   %%%%%%%%%%%
%%%%%%%%%%%%%   Section 2   %%%%%%%%%%%
\section{The No $\bf{a}$-$\rm{p}$ and $\bf{b}$-$\rm{p}$ Lemma}\label{sec:APreAndBPrep}

%%%%%%%%%%%%%   Sub-Section 2.1   %%%%%%%%%%%
%%%%%%%%%%%%%   Sub-Section 2.1   %%%%%%%%%%%
\subsection{Preparation of spin-$\frac{1}{2}$ particles along a vector}\label{sec: Preparation of spin12 particles along a vector}

Consider a $i$-indexed sequence of neutral spin-$\frac{1}{2}$ particles that is prepared along unit vector $\bf{a}$.  One knows that the values $u_i$ of the measurements of the normalized projection of the spin of $p_i$ along $\bf{a}$ would be $u_i\equiv 1$.  Then, for any vector $\bf{\alpha}$ along which one measures the normalized projection $x_i$ of the spin of $p_i$, after the preparation, QM teaches us that:
\begin{equation}\label{Cor0}
\langle u_i\cdot x_i\rangle=\bf{a}\cdot\bf{\alpha},
\end{equation}
which is the spin-$\frac{1}{2}$ form of the well known  \emph{Malus law} that was first formulated, about two centuries ago, for polarizations.
%%%%%%%%%%%%%   Sub-Section 2.2   %%%%%%%%%%%
%%%%%%%%%%%%%   Sub-Section 2.2   %%%%%%%%%%%
\subsection{Preparation of spin-$\frac{1}{2}$ particles along an axis}\label{sec: Preparation of spin12 particles along an axis}
Similarly, if the integers are split into two complementary sets, say $N_{\bf{a}}$ and $N_{-\bf{a}}$, such that:

$\qquad$ - For $i$ in $N_{\bf{a}}$ the preparation is along $\bf{y}_i=\bf{a}$, so that the value of a measurement of the relative value of the normalized projection of the spin of $p_i$ along $\bf{a}$ would be $u_i=1$ for  $i \in N_{\bf{a}}$.

$\qquad$ - For $i$ in $N_{-\bf{a}}$ the preparation is along $\bf{y}_i=-\bf{a}$, so that the value of a measurement of the relative value of the normalized projection of the spin of $p_i$ along $\bf{a}$ would be $u_i=-1$ for  $i \in N_{-\bf{a}}$.

\medskip
\noindent
Then, for any vector $\bf{\alpha}$ along which one measures the relative value of the normalized projection $x_i$ of the spin of $p_i$, after the preparation, one still finds:
\begin{equation}\label{Cor1}
\langle u_i\cdot x_i\rangle=\bf{a}\cdot\bf{\alpha}\,, 
\end{equation}
since, from QM, we know that in (\ref{Cor0}), the signs of  $u_i$ and $x_i$ would change simultaneously.  More precisely:

\medskip
$\qquad$ - When $u_i=1$,  $\bf{y}_i=\bf{a}$ so that $x_i$ is as in (\ref{Cor0}).

\medskip
$\qquad$ - When $u_i=-1$, the value of $x_i$ of the relative value of the normalized projection of the spin of $p_i$ along $-\bf{\alpha}$ would be as the $x_i$ if $u_i$ would be equal to $1$ (by change of orientation).  Thus $u_i$ (along $\bf{y}_i=-\bf{a}$ ) and $x_i$ (along $\bf{\alpha}$) change sign together and together only. 

\bigskip
Instead of mentioning explicitly the partition of the integers, one can as well consider that the preparation is made along a sequence of vectors $\bf{y}_i$ in the set $\lbrace \bf{a},\,-\bf{a}\rbrace$, and this is indeed the point of view that we will adopt. Then:

\medskip
$\qquad$ - The usual preparation corresponds to preparation along some vector $\bf{a}$  or equivalently  along the axis carrying $\bf{a}$ with ${\bf{y}}_i \equiv {\bf{a}}$,

\medskip
\noindent
and one can also say (as another equivalent description) that:

\medskip
$\qquad$ - The usual preparation along $\bf{a}$ corresponds to preparation along the axis of $\bf{a}$ and numerical sequence $u_i\equiv +1$.

\bigskip
More generally, we consider preparation along arbitrary sequences of vectors $\{\bf{y}_i\}$ in the set $\lbrace -\bf{a},\,+\bf{a}\rbrace$, and (once the axis is fixed) one can replace the characterization by the vectors $\bf{y}_i\,\in \,\lbrace \bf{a},\,-\bf{a}\rbrace$ by a characterization by a numerical sequence $\{u_i\}$ with $u_i\,\in\, \lbrace -1,\,+1\rbrace$.

The preparation along $\{\bf{y}_i\}$ in the set $\lbrace \bf{a},\,-\bf{a}\rbrace$ can indeed also be considered as a 
\emph{preparation along the axis determined by the set  $\lbrace \bf{a},\,-\bf{a}\rbrace$}, but for that preparation to be fully specified, we also need to specify the sequence $\{\bf{y}_i\}$ or equivalently, once the axis is specified, the sequence $\{u_i\}$.

%%%%%%%%%%%%%   Sub-Section 2.3   %%%%%%%%%%%
%%%%%%%%%%%%%   Sub-Section 2.3   %%%%%%%%%%%
\subsection{An abstract construct inspired by preparation}\label{sec: A new approach to preparation}

Inspired by these considerations about preparation along an axis (instead of along a vector as usual) we propose a new concept that is abstracted from the way we have discussed preparation along an axis.  This is what we formalize in the following:

\medskip
\noindent
{\bf Definition 1 ($\bf{a}$-p sequences of particles.)} \emph {A sequence of particles $(p_i)_{i\in I}$ is $\bf{a}$-p if there exists a sequence $u=(u_i)_{i\in I}$ with values in $\{-1,+1\}^{I}$ such that for every $\bf{\alpha}\in \bf{R}^3$, 
denoting by $x_i=x_i(\bf{\alpha})$ the projection of the spin of $p_i$ along $\bf{\alpha}$, we have:
\begin{equation}\label{Inner}
  \langle {u},{x}\rangle\,=\,\bf{a}\cdot \bf{\alpha}\ ,
\end{equation}
where $x=(x_i)_{i\in I}$.}

\medskip
\noindent
{\bf Remark.} \emph{The projection  of the spin of $p_i$ along $\bf{\alpha}$ must be performed after the sequence $u=(u_i)_{i\in I}\in \{-1,+1\}^{I}$ has been associated to the sequence of particles  $p=(p_i)_{i\in I}$ with matching indices.  This is a precision that is implicit everywhere one expects it but we make it explicit only in association with Definitions 1 and 2 where this precision is crucial to overcome the need of counterfactuals when using these definitions.  In the same vein, it is also important to realize that many values of $\bf{\alpha}$, say  $\{\bf{\alpha}_k\}_{k\in K}$ for some index set $K$  may tested and for that to make sense, the value $\bf{\alpha}_{k(i)}$ chosen to correspond to $i$ has to be chosen after one has chosen $u_i$ : the reader is invited to check that the \emph{No $a$-p and $b$-p sequences Lemma} would readily be proved false (in Boole's Realm) without this precision on the need to choose $u_i$ before $k(i)$.  We also leave to the reader the choice of how to make the time orderings that we need here unambiguous).}

%%%%%%%%%%%%%   Sub-Section 2.3   %%%%%%%%%%%
%%%%%%%%%%%%%   Sub-Section 2.3   %%%%%%%%%%%
\subsection{A question about $\bf{a}$-p sequences}\label{sec: A question about a p sequences}
The new concept of being $\bf{a}$-p for a sequence of particles, although quite anchored in usual preparation ideas and practice, is abstract enough to consider preparation along $n$ axes at once.  We have examined the simplest case when $n=1$.   In the next to simplest case when $n=2$, we can formulate 
a precise question after isolating a definition:

\medskip
\noindent
{\bf Definition 2 ($\bf{a}$-p and $\bf{b}$-p sequences of particles.)} \emph {Suppose $\bf{a}$ and $\bf{b}$ are two unit vectors in $\bf{R}^3$ carried by distinct axes.  We say that a sequence $\{p_i\}_{i\in \bm I}$ of particle is \emph{$\bf{a}$-p and $\bf{b}$-p} if there exist two sequences $u=(u_i)_{i\in I}$ and $v=(v_i)_{i\in I}$ with values in $\{-1\, ,\, +1\}$ such that for every $\bf{\alpha}$ in $\bf{R} ^3$, denoting by $x_i\,=\,x_i(\bf{\alpha})$ the projection of the spin of $p_i$ along $\bf{\alpha}$, we have:
\begin{equation}\label{both1}
\langle {u}, \,{x(\bf{\alpha})}\rangle  =  \bf{a}\cdot \bf{\alpha}\\
\end{equation}
\begin{equation}\label{both2}
\langle {v},\, {x(\bf{\alpha})}\rangle  =  \bf{b}\cdot \bf{\alpha}
\end{equation}
}

\noindent
{\bf Question (Do $\bf{a}$-p and $\bf{b}$-p sequences of particles exist?)} \emph { More precisely: Suppose $\bf{a}$ and $\bf{b}$ are two unit vectors in $\bf{R}^3$ carried by distinct axes. Can a sequence of spin-$\frac{1}{2}$  neutral particles be both $\bf{a}$-p and $\bf{b}$-p ?}
 
 \medskip
 \noindent
{\bf No $a$-p and $b$-p sequences Lemma.} \emph{Suppose $\bf{a}$ and $\bf{b}$ are two unit vectors in $\bf{R}^3$ carried by distinct axes.  Then no sequence of spin-$\frac{1}{2}$ neutral particles can be both $\bf{a}$-p and $\bf{b}$-p.}
%\end{lemma}

%\begin{proof}
\medskip
 \noindent
{\bf Proof.} 

Suppose not, so that (\ref{both1}) and (\ref{both2}) happen for certain sequences $u,v\in \{-1,+1\}^{I}$ and every $\bf{\alpha}\in \bf{R}^3$. Then, using (\ref{BooleBell1}), we can write:

\begin{equation}\label{eq1}
 {\rm Prob}(u_i=x_i(\bf{\alpha}))\;=\; \frac{1+\bf{a}\cdot\bf{\alpha}}{2} \, ,
\end{equation}
as well as:
\begin{equation}\label{eq2}
 {\rm Prob}(v_i=x_i(\bf{\alpha}))\;=\; \frac{1+\bf{b}\cdot\bf{\alpha}}{2} \, .
\end{equation}

Taking $\bf{\alpha}=\bf{a}$ in both (\ref{eq1}) and (\ref{eq2}), we get:
\begin{equation}\label{eq3}
  {\rm Prob}(u_i=x_i(\bf{\alpha}))\;=\;1\ ,
\end{equation}
and: 
\begin{equation}\label{eq4}
  {\rm Prob}(v_i=x_i(\bf{\alpha}))\;=\;\frac{1+\bf{a}\cdot\bf{b}}{2}\ ,.
\end{equation}
From (\ref{eq3}) and (\ref{eq4}) we deduce that:
\begin{equation}\label{eq5}
 {\rm Prob}(u_i=v_i)\;=\;\frac{1+\bf{a}\cdot\bf{b}}{2}\ .
\end{equation}
Given the relationship (\ref{BooleBell1}) between probabilities and correlations, (\ref{eq5}) tells us that:
\begin{equation}\label{eq5a}
 \langle{u},\,{v}\rangle \;=\;\bf{a}\cdot\bf{b}\ .
\end{equation}
Now let us go back to (\ref{eq1}) and (\ref{eq2}) 
%and choose $\bf{\alpha}$ there so that $\bf{\alpha}\cdot\bf{a}=\bf{\alpha}\cdot\bf{b}$. We 
just to recall that we can re-write (\ref{eq1}) and (\ref{eq2}) as (\ref{both1}):
\smallskip
%\begin{equation}\label{eq6}
$$
 \langle{u},\, {x}\rangle\;=\;\bf{a}\cdot\bf{\alpha}
$$
%\end{equation}
and (\ref{both2})
%\begin{equation}\label{eq7}
\smallskip
 $$
 \langle{u}, \, {v}\rangle\;=\;\bf{b}\cdot\bf{\alpha}\ ,
$$
%\end{equation}
\smallskip
respectively.
%\medskip
\noindent
We pause to point out that with (\ref{eq5a}),  (\ref{both1}), and  (\ref{both2})
we have the Malus law form for all the correlations that we need, but it is thanks to 
the new concept of $\bf{a}$-p and $\bf{b}$-p that we could devise a mean to get 
the Malus Law format of (\ref{eq5a}).

\medskip
Now recall the abstract Boole-Bell inequality (\ref{BooleBell}), that reads 
 \[
 |\langle{f},\, {g}\rangle -\langle{f}, \, {h}\rangle| +  \langle{g}, \, {h}\rangle\;\leq\;1\ ,
\]
%\end{equation}
that is valid for all triples of sequences $f,g,h\in \{-1,+1\}^{I}$ and that can be used in both the Boole Realm and the Bell Realm as we mentioned in the Introduction (but for now we only consider the Boole Realm).

\bigskip
Then we distinguish three cases:

\medskip
$\quad$- Assume that the smallest angle $\theta$ between $\bf{a}$ and $\bf{b}$ is acute. Then, one can choose $\bf{\alpha}$ as the vector orthogonal to $\bf{a}$ (respectively to $\bf{b}$) and such that $\bf{b}$ is in the right angle sector determined by $\bf{\alpha}$ and $\bf{a}$ (respectively $\bf{a}$ is in the right angle sector determined by $\bf{\alpha}$ and $\bf{b}$).  In either case (\ref{BooleBell}) is violated by Quantum Mechanics since as a result of Malus law (\ref{BooleBell}) reduces to 
\begin{equation}
\cos (\theta)+\sin (\theta)<1\,, 
\end{equation}
which is false when $\theta$ is acute.

\medskip
 $\quad$- Assume that the smallest angle $\theta$ between $\bf{a}$ and $\bf{b}$ is obtuse. Then, one can choose $\bf{\alpha}$ as the vector orthogonal to $\bf{a}$ (or $\bf{b}$) and such that $\bf{\alpha}$ is in the smallest sector determined by $\bf{a}$ and $\bf{b}$.  In either case (\ref{BooleBell}) is violated by Quantum Mechanics since as a result of Malus law (\ref{BooleBell}) now reduces to 
\begin{equation}
|\cos (\theta)|+\sin (\theta-\frac{\pi}{2})=\cos (\theta-\frac{\pi}{2})+\sin (\theta-\frac{\pi}{2})<1\,, 
\end{equation}
which is false when $\theta$ is obtuse.

\medskip
$\quad$- Assume that the smallest angle $\theta$ between $\bf{a}$ and $\bf{b}$ is a right angle. Then by choosing  $\bf{\alpha}$ along the bisector of the smallest angle between $\bf{a}$ and $\bf{b}$, one gets that (\ref{BooleBell}) is violated by Quantum Mechanics by observing that as a result of Malus law, (\ref{BooleBell}) then reduces to 
\begin{equation}
\cos (\frac{\pi}{4}  )+\sin (\frac{\pi}{4}  )=\sqrt {2}<1\,, 
\end{equation}
one more flagrant contradiction.

\smallskip
\noindent
This concludes the proof of the No $a$-p and $b$-p sequences Lemma.

We point out that the definition of $a$-p and $b$-p sequences and the proof of the No $a$-p and $b$-p sequences  Lemma respectively allow and require that the further spin projection measurement be performed along a freely chosen vector.  We will not dwell on what happens otherwise in the present note, leaving that for later exposition.

%%%%%%%%%%%%%   Section 3   %%%%%%%%%%%
%%%%%%%%%%%%%   Section 3   %%%%%%%%%%%
%%%%%%%%%%%%%   Section 3   %%%%%%%%%%%
%%%%%%%%%%%%%   Section 3   %%%%%%%%%%%
\section{Some hypotheses for classical and more recent Bell type theories}\label{sec:Hypotheses}

So far, we have been using Quantum Mechanics, staying definitely in Boole' s Realm, but it is now Bell's Realm that we want to explore.  For that we have to augment QM by (1) a Realism hypothesis that may seem contradictory with QM to people  who have a strong conviction about the (generic) lack of Realism at the scale where
QM operates and (2) a second assumption. We examine two augmentations of QM, using the same form of Realism, that is the weakest form of Realism letting one develop \emph{``Bell's Theory"} (\emph{i.e.,} the collection of Bell's type Theorems and the definition of the conditions to get there). For the second assumption there are two choices that we will consider.  One was used by Bell in \cite{Bell} and borrowed from Einstein: it goes under the name of \emph{Locality}. The other is the \emph{Effect After Cause Principle} (EACP) that was introduced in \cite{Tresser1} and refined in \cite{Tresser2} and that we will recall here as well as all the main elements of Bell's Realm.

\medskip
The following two conventions are adopted in a more or less explicit form in all works on Bell's Theory, independently of the strength of the Augmentation being chosen:

\bigskip
\noindent
\textbf{Convention 1.}  \emph{Whenever we assume that Quantum Mechanics is augmented by a form of Realism, we implicitly postulate that \emph{any quantity that is not measured but that exists according to the Augmentation has the value that would have been measured if this quantity would have been the one measured, the World being otherwise unchanged}.}

\medskip
\noindent
\textbf{Convention 2.}  \emph{Whenever we assume that Quantum Mechanics is augmented by a form of Realism, we assume that said Augmentation is made \emph{without changing the statistical predictions}.  This is (up to wording) the assumption that Bell made in his foundational 1964 paper \cite{Bell}, except for the fact that we do not restrict the choice of Augmentation to Predictive Hidden Variables.}

\medskip
\noindent
\textbf{Definition 3 (Locality).}  \emph{ \emph{Locality} tells us that if $(x_0, t_0)$ and $(x_1, t_1)$ are spatially separated, \textit{i.e.,}  $\Delta x ^2> c^2 \Delta t^2$, with $c$ standing for the speed of light,  then  the output of a measurement made at  $(x_1, t_1)$ cannot depend upon the setting of an instrument at  $(x_0, t_0)$ (making no measurement at $(x_0, t_0)$ being in this context one possible apparatus setting). Furthermore, if one assumes that Weak Realism as defined below holds true,  the value of any observable that could be measured at $(x_1, t_1)$ in lieu of the observable that is actually being measured there is also independent of any instrument setting at $(x_0, t_0)$.}

\medskip
\noindent
\textbf{Definition 4 (Weak Realism).} \emph{\emph{Weak Realism} (or \emph{WR}) postulates that here is a value associated to any {\bf useful} unperformed measurement that could have been made on a particle at the time when some other measurement was actually made instead on that particle.  Here an unperformed measurement
is deemed useful if the (generally unknown) value for the corresponding observable, that could exist by invoking the chosen form of Realism, is used to formulate a Bell's Theorem. Furthermore, it is assumed that the values of the observable that exist according to Weak Realism preserve the statistical predictions of Quantum Mechanics.}

The usefulness prescription (whereby for instance at most one extra spin projection is assumed to exist per particle for which one spin projection is actualy measured) makes WR slightly weaker than Stapp's  \emph{Contrafactual Definiteness} \cite{Stapp1985} (see also \cite{Stapp1971}), also called \emph{Macroscopic Counterfactual Definiteness} and abbreviated as \emph{MCFD} by Leggett who offers an interesting discussion of it in \cite{Leggett2008}.  MCFD is not only implied by the hypothesis used by Bell in \cite{Bell} but also by Microscopic Realism as explained in \cite{Leggett2008} and also by what is called sometimes \emph{the EPR  condition of reality} \cite{EPR}.  

\medskip
Thus the Augmentation of Quantum Mechanics that we choose in order to develop a restricted hypotheses version of Bell's Theory is``Weak Realism", the small modification of MCFD whereby we have have added the word {\bf ``useful"} to the definition of MCFD.  We have added that word so that WR be thus somewhat explicitly constructed such as it is:

\smallskip
- \emph{The weakest form of Realism sufficient to develop Bell's Theory}  

\smallskip
\noindent
while 

\smallskip
- \emph{Preserving the statistical predictions of Quantum Mechanics (by a standard important extra hypothesis going back to\cite{Bell}: see also Convention 2 above).}

\noindent
This is why we call the chosen Augmentation simply \emph{Weak Realism} (``Weakest Realism" sounding  less appealing).  We will say \emph{``Realism"} to mean \emph{``any form of Realism"}, and whenever \emph{``Weak Realism"} is invoked,  \emph{``any (reasonable) form of Realism"} could be used instead.

\medskip
\noindent
\textbf{Effect After Cause Principle (\emph{EACP} - General Form):} \emph{- (i) For any Lorentz observer the measured value of an observable (once measured) cannot change as a result of any cause that happens after said observable has been measured for that observer.}

\noindent
\emph{- (ii) Furthermore, assume now that Weak Realism holds true and that an observable is measured at the point  $(x,t)$ of a Lorentz observer. Then consider another observable that gets a definite, albeit generally unknown, value at  $(x,t)$ for the Lorentz observer because of Weak Realism. In parallel to what is described in (i) for a measured value, the definite value of the second observable at $(x,t)$ cannot change as a result of any cause that happens after said non-observed observable has been assigned, as a result of Weak Realism, a definite value at the point  $(x,t)$ of that observer.}

\bigskip
We will mostly use the EACP in a form that is much more specific that the one proposed here (see after the following Warnings) but we have judged that it was preferable to stay at this less technical level to begin with.  The subtlety of (both of) the statement(s) of the EACP calls for some special comments that are more of the ``warning" type than ordinary remarks.

\medskip 
\noindent
\textbf{Warnings.} \emph{
\begin{description}
\item[W1] We notice that time ordering used in the definition of the EACP is relative to the chosen Lorentz observer, as it should be. 
\item[W2] Accepting that Locality fails to always hold true means that the value of an observable may well depend on a ``later" event in the time ordering of the chosen Lorentz observer: see Warning [W1]. The next warning [W3] provides limits for the dependence on later events. 
\item[W3] A reading of an observable, once performed (or potentially performed when dealing with entities that only exist if one assumes Weak Realism) cannot (further) change because of a later cause, but the reading may \emph{depend} or \emph{not depend} on a cause that happens later for some Lorentz observer according to whether one assumes Non-Locality or Locality.
\end{description}
} 

%\bigskip
\smallskip
\noindent
\textbf{Effect After Cause Principle (\emph{EACP: Specific Form}).} \emph{For any Lorentz observer and for any $\mathcal Q$ in the set of effective (unprimed) and Realism dependent (primed) sequences of observable values $\{\mathcal E,\, \mathcal E',\, \mathcal P ,\, \mathcal P'  \}$, a value $\mathcal Q_i$ of $\mathcal Q$ cannot change as a result of a cause that happens after $\mathcal Q_i$ has been attributed to $\mathcal Q$ for said Lorentz observer, both when $\mathcal Q_i$ is measured and when the value $\mathcal Q_i$ (then generally unknown) is attributed with a definite value to $\mathcal Q$ by Weak Realism.}

\medskip
This version of the EACP adapted to the context of Bell's Theory will be used to prove the following two comparison results that are crucial to our purpose regarding Bell's Theory.

\medskip
We recall the following result from \cite{Tresser1} and \cite{Tresser2}; it responds by the negative to many questions (in fact always essentially the same question but asked in different forms by many people) raised about whether the EACP is or not Locality under disguise.

\medskip
\noindent
\textbf{EACP vs Locality Lemma.} \emph{The EACP is different from Locality and not stronger than Locality, meaning that the EACP does not imply Locality.}

\smallskip
\noindent
\emph{Proof of  the EACP vs Locality Lemma.} We assume Weak Realism and the EACP and notice that, in view of Warning [W1]-[W3], the EACP has been formulated so as to be compatible either with Locality or with Non-Locality, whichever one chooses.  \textbf{Q.E.D.} 

\medskip
\noindent
\textbf{EACP weaker than Locality.} \emph{ The EACP is indeed weaker than Locality. Otherwise speaking, Locality implies the EACP but the reverse implication is not true.}

\medskip
\noindent	
\emph{About the EACP weaker than Locality statement.}  The relative strength statement follows from the easily checked fact that the failure of the EACP permits \emph{Superluminal Message Transmission} (or SMT), using readable messages (the ones we all know how to use) or at least (depending on how the EACP fails to hold true) \emph{Realist Superluminal Message Transmission} (or RSMT), \textit{i.e.,} a form of SMT using messages that are collections of values some or all of which only exist because of the chosen Realism assumption.  Since we know that the violation of Locality does not permit SMT (see, \textit{e.g.,} \cite{Eberhard1978},  \cite{GRW1980}, \cite{Jordan1983}), we know that Locality and \emph{Causality} (the impossibility of SMT) do not coincide in a World without augmentation of Quantum Mechanics by any form of Realism and that indeed Locality is stronger that Causality  in such a World. Furthermore Locality is stronger than Causality extended to be the impossibility of both SMT and RSMT in any World when assume the impossibility of RSMT as we do since RSMT takes care of the case when the negation
of EACP is by allowing RSMT, even if SMT is not permitted.  We deduce that the EACP is weaker than Locality in a World without augmentation of Quantum Mechanics by any form of Realism, and more generally as long  (at least) as one considers RSMT to be impossible.  We do consider RSMT to be as impossible as SMT, independently of whether Weak Realism is compatible with the actual laws of Physics or not, on the basis that \emph{an extension of Physics by variables that are not only inaccessible but also violate fundamental laws obeyed by regular variable basically escapes the Realm of Physics}.  Hidden Variables, one of the strong forms of Microscopic Realism, are unaccessible but otherwise obey the law of Physics, and many who are ready to support Non Locality rely for that on the fact that Non Locality does not permit SMT: it does not seem compatible with Physics to allow RSMT on the sole basis that it is all about things out of our reach. Details are left to the reader.  
\section{Bell's Theorem in a nutshell}%%%%QQQQQQQQQQQ
%%%%%%
%%%%%%%%%%%%%%%%%%%%%%%%%%%%%%%
%%%%%%%%%%%%%%%%%%%%%%%%%%%%%%%%%%%%%
%%%%%%%%%%%%%%%%%%%%%%%%%%%%%%%%%%%%%
%\medskip
We now recall that in the founding paper of Bell's Theory \cite{Bell}, by showing that (\ref{BooleBell}) hold in Bell's Realm and that QM violates (\ref{BooleBell}) Bell (see page 199 of that paper) reached the conclusion that:

\smallskip
\emph{``In  a theory in which parameters are added to quantum mechanics to determine the results of individual measurements, without changing the statistical predictions, there must be a mechanism whereby the setting of one measuring device can influence the reading of another instrument, however remote.  Moreover, the signal involved must propagate instantaneously, so that such a theory could not be Lorentz invariant."}.

\bigskip
\noindent
More generally, the structure of a typical Bell type theorem reads either as the following statement that we call \emph{the Main Implication} or as its consequences as in Bell's citation just above:

\medskip
\noindent
\textit{ \,Quantum Mechanics} \quad \quad \quad \textit{Some inequality is violated }

\noindent
\textit{+ Augmentation choice} \,\,\,$ \Rightarrow \,\,$\textit{ for appropriate choices} 

\noindent
\textit{+ Extra hypothesis}\qquad \qquad \, \textit{ of some parameters.}

\medskip
\noindent
In the terms of the Main Implication, the example of ``Augmentation" chosen in Bell's 1964 paper \cite{Bell} is the assumption that there are ``Predictive Hidden Variables with the same statistics as Quantum Mechanics" while Bell's original example of ``Extra hypothesis" is ``Locality" that we next redefine both more formally and in such a way that the role of the augmentation be clearly stated. In fact, while Bell uses    
``Predictive Hidden Variables with the same statistics as Quantum Mechanics" in order to derive his theorem, he explain how to deduce these variable using Bohm's 
explanations of the EPR paper (in all rigor, Bell mentions a paper of Bohm and Aharonov  \cite{BohmAharonov1957} where the Bohm version \cite{Bohm} in terms of  spin-$\frac{1}{2}$  particles is recalled and then recast in terms of photons and their polarizations in order to propose an experimental verification of the EPR paper).
Some people have tried to reduce the set of hypotheses used by Bell to [QM + Locality] but we will not get into a dispute of such claims (that, we consider, overlook a realist interpretation going back to \cite{Bohm}) since anyway we can consider as hypotheses (thus possibly deduced from more primitive ones) leading to a Bell's Theorem the following augmentations  whose elements (except for Quantum Mechanics here appearing as QM) have just been recalled:

\bigskip
$\quad$ - Hypothesis 1: \emph{[ QM+ Weak Realism+ Locality]}.   

\medskip
$\quad$ - Hypothesis 2: \emph{[ QM+ Weak Realism+ EACP]}.   

As we shall see next, both hypotheses that let us get into Bell's Realm and from which one has deduced the corresponding Bell's Theorem, also let us predict that a particle, say $p$ (out of a pair $(p,\,p')$) in the singlet state must be both $\bf{a}$-p and  $\bf{b}$-p where  $(-\bf{a})$  is the vector along which the normalized spin of $p'$ is measured and  $(-\bf{b})$  is the vector along which the projection f the spin of $p'$ is assumed to make sense when assuming Weak Realism.

%%%%%%%%%%%%%   Section 4   %%%%%%%%%%%
%%%%%%%%%%%%%   Section 4   %%%%%%%%%%%
%%%%%%%%%%%%%   Section 4   %%%%%%%%%%%
%%%%%%%%%%%%%   Section 4   %%%%%%%%%%%
\section{Applications of the No $\bf{a}$-$\rm{p}$ and $\bf{b}$-$\rm{p}$ Lemma}
It is plain that under Hypothesis 1:

- One of the particles, say $p$ (out of a pair $(p,\,p')$) in the singlet state must be both $\bf{b}$-p and  $\bf{b'}$-p where  $(-\bf{b})$  is the vector along which the normalized spin of $p'$ is measured and  $(-\bf{b'})$  is the vector along which the projection f the spin of $p'$ is assumed to make sense when assuming Weak Realism.

- The other particle $p'$ (out of the pair $(p,\,p')$) in the singlet state must be both $\bf{a}$-p and  $\bf{a'}$-p where  $(-\bf{a})$  is the vector along which the normalized spin of $p$ is measured and  $(-\bf{a'})$  is the vector along which the projection of the spin of $p'$ is assumed to make sense when assuming Weak Realism.

Thus Hypothesis 1 leads to a contradiction if one considers entanglements as used to develop Bell's theory.

As for Hypothesis 2, one generally considers the use of Weak Realism only on one of the particle, say $p'$ (only then can a Bell inequality be proven to hold: see \cite{Tresser1}, \cite{Tresser2}). Then for a Lorentz observer for whom the data from $p$ (obtained by Alice) are anterior to those about  $p'$ and obtained by Bob, the data about $p$ cannot be affected by non-locality since those data are anterior to those about $p'$.  Thus, assuming the EACP, non-local effect cannot happen at $p$.  But for that Lorentz observer, or anyone indeed, the data at $p'$ cannot be affected by non-locality either since on $p$, only one spin measurement is made and no spin value is used nor even assumed to make sense by invoking Weak Realism on particle $p$. Thus we are back to assuming Hypothesis 1 by isolating 
a case where the EACP has the same effect as Locality 
despite the general status of EACP vs Locality described above.

Thus we have proved the following:

 \medskip
 \noindent
{\bf  Theorem: Boole inequalities prevent Bell's inequalities from having physical meaning.} \emph{At least one of Bell's inequalities has no physical content, in the sense of implying a contradiction by conjunction of the No $a$-p and $b$-p sequences Lemma that only assumes Quantum Mechanics and one of Hypothesis 1 or Hypothesis 2.}

\medskip
We have also tried to ignore Convention 1 and used non-conventional possible meaning of the values that would be taken by observables assuming weak Realism: it turns out than none of these tricks works.

\medskip
The results presented here may shed new light on results recently presented in \cite{PuseyBarrettRudolph} and \cite{HallOnPuseyBarrettRudolph} that have been pointed to our attention while we were editing the present paper.
\bigskip

\bigskip
\begin{acknowledgments}
\textbf{Acknowledgments:} 
These results were presented in preliminary form by CT at the 90$^{\rm th}$ Birthday Party of Mauricio Peixoto April 13 - April 15, 2011 at I.M.P.A. in Rio de Janeiro, and the 70$^{\rm th}$ Birthday Party of Dennis Sullivan at the Simons Center for Geometry and Physics in Stony Brook May 26 - June 4, 2011.  CT thanks Franck Lalo{\"e} for calling his attention on \cite{PuseyBarrettRudolph} and \cite{HallOnPuseyBarrettRudolph}  just before the authors finished editing this paper never  discussed with Franck before submission to [quant-ph] since CT was discussing other matters with him. EdF received support from FAPESP through "Projeto Tematico Dinamica em Baixas Dimensoes", Proc. FAPESP 2006/03829-2. 
\end{acknowledgments}


\begin{thebibliography}{99}
\bibitem{Bohm}
D.\ Bohm \textit{Quantum Theory,} (Prentice Hall; New York 1951).

\bibitem{EPR}
A.\ Einstein, B.\ Podolsky, N.\ Rosen \textit{Phys. Rev.} \textbf{47}, 777 (1935).

\bibitem{Jammer74}
M.\ Jammer, \textit{The Philosophy of Quantum Mechanics: The
Interpretations of Quantum Mechanics in Historical Perspective,}
(John Wiley \& Sons Inc; New York, 1974).

\bibitem{Jammer85}
M.\ Jammer, in \textit{Symposium on the Foundations of Modern Physics: 50 years of the Einstein-Podolsky-Rosen Gedankenexperiment,}  P. Lahti and P. Mittelstaedt, Eds.  (World Scientific, Singapore, 1985), pp. 129-149. 

\bibitem{FineShaky}
A.\ Fine, \textit{The Shaky Game; Einstein Realism and the Quantum Theory,}
(The University of Chicago Press; Chicago, 2$^{nd}$ edition, 1996).

\bibitem{EinsteinIdeasAndOpinions}
A.\ Einstein, \textit{Ideas and Opinions,} (Crown Publishers; New York, 1954).

\bibitem{Schilpp}
P.A.\ Schilpp, (editor) \textit{Albert Einstein:
Philosopher-Scientist} (The Open Court Publishing Co.; La Salle,
IL, 3$^{\rm rd}$ edition, 1969 (Vol. 1) - 1970 (Vol. 2)).

\bibitem{Stapp1971} 
H.P.\ Stapp, \textit{Phys. Rev.} \textbf{3 D}, 1303 (1971).

\bibitem{Bell}
J.S.\ Bell, \textit{Physic} (Long Island City, NY)  \textbf{1}, 195 (1964).

\bibitem{Boole1854}
G.\ Boole, \textit{An Investigation of the Laws of Thought,} (Walton \& Maberly; London, 1854); (Dover; New York, 1958).

\bibitem{Boole1862}
G.\ Boole, \textit{Phil. Trans. Roy. Soc. (London)}  \textbf{152}, 225 (1862).

\bibitem{Pitowsky2001}
I.\ Pitowsky, \textit{Phys. Rev.  A} \textbf{64}, 4102 (2001).

\bibitem{Tresser1}
C.\ Tresser, \textit{Eur. Phys. J.}  \textbf{D 58}, 385 (2010).

\bibitem{Tresser2}
C.\ Tresser, \textit{ Eur. Phys J.} {\bf D 62}, 139 (2011).  

\bibitem{Stapp1985}
H.P.\ Stapp, in \textit{Symposium on the Foundations of Modern Physics: 50 years of the Einstein-Podolsky-Rosen Gedankenexperiment,}  P. Lahti and P. Mittelstaedt, Eds.  (World Scientific, Singapore, 1985), pp. 637-652. 

\bibitem{Leggett2008}
A.J.\ Leggett, \textit{Rep. Prog. Phys.} \textbf{71}, 022001 (2008).

\bibitem{BohmAharonov1957}
D.\ Bohm, Y.\ Aharonov \textit{Phys. Rev.} \textbf{108}, 1070 (1957).

\bibitem{Eberhard1978}
P.\ Eberhard, \textit{Nuovo Cimento} \textbf{46B} , 392 (1978). 

\bibitem{GRW1980}
G.C.\ Ghirardi, A.\ Rimini, and T.\ Weber, T., Lettere Al Nuovo Cimento \textbf{27}, 293 (1980).

\bibitem{Jordan1983}
T.\ Jordan, \textit{Phys. Lett. A} \textbf{94}, 264 (1983).

\bibitem{PuseyBarrettRudolph}
M.F.\ Pusey, J.\ Barrett, and T.\ Rudolph, \emph{The quantum state cannot be interpreted 
statistically,} arXiv:1111.3328v1 [quant-ph] 14 Nov 2011.

\bibitem{HallOnPuseyBarrettRudolph}
M.J.W.\ Hall, \emph{Generalisations of the recent Pusey-Barrett-Rudolph theorem for statistical models of quantum phenomena,} arXiv:1111.6304v1 [quant-ph] 27 Nov 2011

%\bibitem{Bohr}
%N.\ Bohr, \textit{Phys. Rev.} \textbf{48}, 696 (1935).
%
%%%%%%%%%%\bibitem{Tresser3}
%%%%%%%%%%C.\ Tresser \textit{EPR and the Einstein-Tolman-Podolsky paper: any is not all in microphysics}, modified from quant-ph/0503006.
%%%%%%%%%%
%%%%%%%%%%\bibitem{ETP}
%%%%%%%%%%A.\ Einstein, R.C.\ Tolman, B.\ Podolsky, \textit{Phys. Rev.} \textbf{37}, 780 (1931).
%%%%%%%%%%
%%%%%%%%%%\bibitem{CHSH69}
%%%%%%%%%%J.F.\ Clauser, M.A.\ Horne, A.\ Shimony, R.A.\ Holt, \textit{Phys.
%%%%%%%%%%Rev. Lett.} \textbf{23}, 880 (1969).

\end{thebibliography}
%
%

\end{document}